\documentclass[a4paper,12pt,twoside]{article}

\usepackage[usenames]{color}

\usepackage{titlesec}

\usepackage[toc,page]{appendix}

\usepackage[margin=1.0in]{geometry}
\usepackage{placeins}
\usepackage{mathtools}
\usepackage{amsmath,amssymb,amsfonts}
\usepackage{amsthm}%theroems proof

\renewenvironment{proof}{{\bfseries Proof:}}{\qed\\}

\usepackage[utf8]{inputenc}
\usepackage[T1]{fontenc}
\usepackage{algorithmic}
\usepackage{algorithm}

\usepackage{textcomp}
\usepackage{nccmath}
\usepackage{color,soul}%highlighting

\usepackage{graphicx}%charge les .eps .For caption options

\usepackage{caption} %pour fig
\usepackage{subcaption} %pour multifigs

\usepackage{tikz-cd}

\usepackage{array}

\makeatletter
\newcommand{\thickhline}{%
    \noalign {\ifnum 0=`}\fi \hrule height 1pt
    \futurelet \reserved@a \@xhline
}
\newcolumntype{"}{@{\hskip\tabcolsep\vrule width 1pt\hskip\tabcolsep}}
\newcolumntype{|}{@{\hskip\tabcolsep\vrule width 0.5pt\hskip\tabcolsep}}
\usepackage{epstopdf}
\usepackage{microtype}
\usepackage{booktabs} % for professional tables
\usepackage{makecell}
\usepackage{pifont}% http://ctan.org/pkg/pifont
\usepackage{footnote}
\makesavenoteenv{tabular}
\makesavenoteenv{table}
\usepackage{footnotebackref}
\usepackage{scrextend}
\usepackage{float}
\usepackage[intoc]{nomencl}
\usepackage{etoolbox}
\makenomenclature
\usepackage{etoolbox}
\renewcommand\nomgroup[1]{%
  \item[\bfseries
  \ifstrequal{#1}{A}{Abbreviations}{%
  \ifstrequal{#1}{O}{Mathematical operators}{%
  \ifstrequal{#1}{S}{Symbols}{}}}%
]}

\usepackage{todonotes}
\usepackage{hyperref}%Last package is href
\hypersetup{
    colorlinks=true,
    linkcolor=blue,
    filecolor=magenta,      
    urlcolor=cyan,
    citecolor=red,
}

\newcommand{\tr}[1]{\mathop{\kern0pt tr} \!\left(#1\right)}

\newcommand{\fx}[2]{\mathop{\kern0pt #1}\!{\left(#2\right)}}%General function

\newcommand{\prt}[1]{\left(#1\right)}%in brackets
\newcommand{\bd}[1]{\boldsymbol{#1}}%in brackets
\newcommand{\gui}[1]{``#1''}%guillemets

\def\BibTeX{{\rm B\kern-.05em{\sc i\kern-.025em b}\kern-.08em
    T\kern-.1667em\lower.7ex\hbox{E}\kern-.125emX}}

\usepackage{fancyhdr}

 \fancypagestyle{plain}{%
\fancyhf{}
\fancyfoot[RO]{\makebox[0pt][l]{\makebox[0cm][r]{\thepage}}}%
\fancyfoot[RE]{\makebox[0pt][r]{\makebox[0cm][l]{\thepage}}}%
}

\begin{document}

\title{A Curious Link Between Prime Numbers, the Maundy Cake Problem and Parallel Sorting.}

% ############################################################################
% Author(s) (no blank lines !)
\author{
% ############################################################################
Jonathan Blanchette, Robert Laganière\\
School of Electrical Engineering and Computer Science\\
University of Ottawa\\
Ottawa, Canada K1N 6N5\\
{\em jblan016@uottawa.ca}
% ############################################################################
} % end-authors
% ############################################################################

\maketitle
\begin{abstract}
   We present new theoretical algorithms that sums the n-ary comparators output in order to get the permutation indices in order to sort a sequence. By analysing the parallel ranking algorithm, we found that the special comparators number of elements it processes divide the number of elements to be sorted. Using the divide and conquer method, we can express the sorting problem into summing output of comparators taking a prime number of elements, given that this prime number divides the initial disordered sequence length. The number of sums is directly related to the Maundy cake problem. Furthermore, we provide a new sequence that counts the number of comparators used in the algorithms.
\end{abstract}

\section{Introduction} \label{sec:intro}
% ############################################################################
\par{Computing is generally done using binary comparators. Perhaps the most intuitive way of sorting is by using brute force. If we count the number of times an element is larger than another, then effectively we indirectly are implementing Insertion Sort or the similar Selection sort. By counting the number of times an element is larger than the others in the sequence, we effectively ranked the element. By obtaining the rank, we can sort the sequence with a permutation. However, ranking (to count the number of times an element is larger than another) can be assigned to multiple processors simultaneously and is therefore highly parallelisable. In the insertion sort or the selection sort algorithm, the depth is of the order of the sequence length. Since we rank in parallel, we have in the end a reduction (or a sum) of $\mathcal{O}\prt{N}$ comparator arrays. Again, using GPU's, this can be reduced in $\mathcal{O}\prt{\log\prt{N}}$ time. But in most algorithms, comparators used are binary. In this paper, we study if we can use comparators that take more than 2 numbers. More specifically, we see what comparator sizes are possible given $N$, the size of the sequence to be ranked. When we count the number the number of partial ranks(output of a comparator bank) required to be reduced, we get a direct connection to the Maundy cake problem \cite{berlekampwinning}. Additionally, we analyse the number of comparators that will have to be used and we provide a new number theoretical sequence related to them.} \par{The new divide and conquer algorithms are mainly theoretical since it is not scalable for large $N$. It still requires $\mathcal{O}\prt{N^2}$ processors running in parallel in the worse case and it requires to factorise a large number which is a nondeterministic polynomial time (NP) problem \cite{krantz2011proof}. Another possible drawback is that the processors need to output $n$-ary comparators, hence they could need special hardware to implement them.}
\par{
We present in this section a way of dividing the parallel selection sorting problem into prime comparators and their unique connection patterns.

Sorting a sequence is a non-linear in the sense that the matrix that transforms the sequence is a permutation matrix $\boldsymbol{P}_{\boldsymbol{x}}$ that depends on the data itself, i.e. 
\begin{equation}\label{eq:sort}
 \boldsymbol{s}=\boldsymbol{P}_{\boldsymbol{x}}^T\boldsymbol{x}
 \end{equation}
   We will drop the data dependence subscript for brevity so $\boldsymbol{P}_{\boldsymbol{x}}=\boldsymbol{P}$. Furthermore,
\begin{equation}\label{eq:permmat}
\boldsymbol{P}=\begin{bmatrix}
         \boldsymbol{e}_{\pi_0} &   \hdots & \boldsymbol{e}_{\pi_{N-1}}
        \end{bmatrix}
\end{equation}
where $\boldsymbol{e}_i$ is a zero $N\times1$ vector except the $i$th entry equals 1. Hence if we define $\boldsymbol{n}_{N\times1}$ to be:
\begin{equation}
\boldsymbol{n}_{N\times1}=\begin{bmatrix}
         0 & 1 &  \hdots & N-1
        \end{bmatrix}^{T}
\end{equation}
        then, 
        \begin{equation}\label{eq:permmat2}
\boldsymbol{P}^T\boldsymbol{n}=\boldsymbol{\pi}
\end{equation}
where $\boldsymbol{\pi}=\begin{bmatrix}
         \pi_0 & \hdots & \pi_{N-1}
        \end{bmatrix}^{T}$.
        }
\section{Converting a sorting problem into a sum problem.} \label{sec:parallelrankfilt}
 Note that, unless otherwise stated, Zero-indexing is used in this project. 
 If we consider equation \ref{eq:sort} and \ref{eq:permmat} we see that what really is important is the indices $\bd{\pi}$ to order a sequence. That is if we solve for $\bd{\pi}$, we indirectly know what our sorted sequence $\bd{s}$ is since $\bd{\pi}$ fully characterizes the permutation matrix. In this section we will use comparator units that output ranks or keys $\bd{\pi}$ of the corresponding input sequence number $\bd{x}$, instead of using them to output a \gui{swapped in order} sequence $\bd{s}$. If we consider that a K-ary comparator $\mathcal{C}_{K\times1}$ or simply $\mathcal{C}_{K}$, it would output keys $\in[{0,1,\cdots,K-1}]$:
\begin{equation}\label{eq:comparator}
\mathcal{C}_{N}\left(\bd{x}_{N\times1}\right)=\bd{\pi}
 \end{equation}
  For example, $\mathcal{C}_{3}\left(\begin{bmatrix}
         6.4 & -9.3 & 0.1
        \end{bmatrix}^{T}\right)=\begin{bmatrix}
         2 & 0 & 1
        \end{bmatrix}^{T}$
        and
        $\mathcal{C}_{2}\left(\begin{bmatrix}
         -40.56 & 10.76
        \end{bmatrix}^{T}\right)=\begin{bmatrix}
         0 & 1
        \end{bmatrix}^{T}$.
        
        Consider for now that the sequence has no repeated elements. Then, for the case where we have exactly 2 elements to compare, we can model $\mathcal{C}_{2}$ as being a boolean function that outputs $0$ and $1$ for false and true respectively:
     \begin{equation}\label{eq:comparatorbinary}
\mathcal{C}_{2}\left(\begin{bmatrix} 
    x_0 \\
    x_1 \\
    \end{bmatrix}\right)=\begin{bmatrix} 
    \left(x_0>x_1\right) \\
    \left(x_1>x_0\right) \\
    \end{bmatrix}=\begin{bmatrix} 
    \delta\left[x_0-x_1\right] \\
    \delta\left[x_1-x_0\right] \\
    \end{bmatrix}=\bd{\pi}
 \end{equation}  
 The relation above is important enough to be given a special notation $c_{i,j}$ for output of the comparison $\left(x_i>x_j\right)=\delta[x_i-x_j]=c_{i,j}$. We can rewrite eq. \ref{eq:comparatorbinary} into:
 
 \begin{equation}\label{eq:comparatorbinary2}
\mathcal{C}_{2}\left(\begin{bmatrix} 
    x_0 \\
    x_1 \\
    \end{bmatrix}\right)=\begin{bmatrix} 
    c_{0,1} \\
    c_{1,0} \\
    \end{bmatrix}
 \end{equation} 
\subsection{Binary partial ranks} \label{subsec:binary}
The first algorithm is very intuitive. To illustrate it with an example, if in a sequence of 16 numbers I told you that one of the elements is bigger than exactly 6 elements, then we know that this is the seventh element. We don't need to know the order of the other elements to infer this. Hence it is possible to do sorting in parallel by finding the permutation index of a sequence element by counting the number of times one element is higher that the other. So we sum together the partial ranks to get the correct index. In vector notation, the permutation vector $\bd{\pi}_{N\times 1}$ is a sum of elements of partial ranks stored in a matrix $\bd{C}_{N\times N}\left(\bd{x}\right)$ or just $\bd{C}_{N}$ .
\begin{equation}\label{eq:permutationvector}
  \mathcal{C}_{N}\left(\bd{x}\right)=\bd{\pi}=\bd{C}_{N} \bd{1}_{N\times 1}
\end{equation}
The entry $(i,j)$ of matrix $\bd{C}_{N}$ are defined by:
%\begin{equation}\label{eq:cs}
\[
c_{i,j}= 
\begin{cases}
    \delta[x_i-x_j],& \text{if } i>j\\ 1-\delta[x_i-x_j],& \text{if } j>i\\ 0              & \text{if } i=j
\end{cases}
\]
%\end{equation}
or perhaps more clearly:
\begin{equation}\label{eq:bigC}
\bd{C}_{N} = \begin{bmatrix} 
    0 & c_{0,1} & c_{0,2} & \dots \\
    \bar{c}_{0,1} & 0 &c_{1,2}& \dots \\
    \bar{c}_{0,2} & \bar{c}_{1,2} & 0 & \dots \\
    \vdots & & & \ddots  \\
    \end{bmatrix}
\end{equation}
Here $\bar{c}$ is the 1-bit complement of $c$. The diagonal is assigned a $0$ by default because it never contributes to the ordering permutation. The lower diagonal are filled with complements of the upper diagonal elements because this way the permutation indices can't repeat themselves. Note that if $\bd{x}$ has repeated values there could be multiple possible solution for $\bd{\pi}$ since the elements $\pi_i$ could be repeated. Despite that, even if there are equal elements in the vector $\bd{x}$, the permutation indices will correspond to a sorted vector, i.e. the sorting algorithm is stable. 
The matrix $\bd{C}_{N}$ is skew-symmetric in the 1-bit sense, in other words the negative sign is replaced with the boolean complement. Let us denote the space of the boolean skew-symmetric matrices by $\mathbb{S}^{N\times N}$ and the possible comparison matrix space by $\mathbb{C}^{N\times N}$. In general, not all boolean skew-symmetric matrices correspond to a possible $\bd{x}$, or $\mathbb{C}^{N\times N}\subsetneq \mathbb{S}^{N\times N}$ for $N>2$. For example, there is no sequence $\bd{x}$ that correspond to the following matrix:
\[
\bd{S}_4=\begin{bmatrix} 
    0 & 0 & 0 & 1 \\
    1 & 0 & 1 & 0 \\
    1 & 0 & 0 & 1 \\
    0 & 1 & 0 & 0  \\
    \end{bmatrix}    
 \]
The indexing vector $\bd{\pi}$ will correspond to the indices of a stable sorting algorithm if the permutation indices are unique.
The structure imposed on $\bd{C}_{N}$  also ensures that the solution $\bd{\pi}$ has unique elements. 

\begin{proof}
Let $v\left(\bd{C}_{N}\right)$ denote the $\frac{1}{2}\left(N-1\right)N\times1$ vector obtained from the vectorized matrix $vec\left(\bd{C}_{N}\right)$ by eliminating all the diagonal and lower diagonal elements of $\bd{C}_{N}$. For example if $N=3$,
$\bd{c}=v\left(\bd{C}_{N}\right)=\begin{bmatrix}
         c_{0,1} &   c_{0,2} & c_{1,2}
        \end{bmatrix}^{T}$.
        Then the sum of partial indices is:
        \begin{equation}\label{eq:vectorisedC}
        \bd{C}_{N}\bd{1}_{N\times1}=\bd{\Delta}_{N\times\frac{1}{2}\left(N-1\right)N}\bd{c}+\bd{n}
        \end{equation}
        
        The matrix $\bd{\Delta}_{N\times\frac{1}{2}\left(N-1\right)N}=\bd{\Delta}_{N}$ has the following recursive structure:
	\begin{equation}
	\bd{\Delta}_{N}=\left[
\begin{array}{cc}
\bd{\Delta}_{N-1} & \bd{I}_{N-1\times N-1} \\
\bd{0}_{N-1\times 1}^T & -\bd{1}_{N-1\times 1}^T
\end{array}
\right]
    \end{equation}	 
          The smallest of theses matrices is: $\bd{\Delta}_{2}=\begin{bmatrix}
         1 \\ -1 
        \end{bmatrix}$ . We have\footnote{This is a matrix rank, not to be confused with a sorting order rank.} $rank\left(\bd{\Delta}_{N}\right)=N-1$ and 
        \begin{equation}\label{eq:Dvanishes}
	\bd{1}^T\bd{\Delta}_{N}=\bd{0}_{\frac{1}{2}\left(N-1\right)N\times 1}^T
    \end{equation}	
    Combining equations \ref{eq:permutationvector},\ref{eq:vectorisedC} and\ref{eq:permmat2} we obtain:
    \begin{equation}\label{eq:C2Pn}
        \bd{\Delta}_{N}\bd{c}+\bd{n}=\bd{P}^T\bd{n}
        \end{equation}
        Assume that indeed $\bd{\pi}$ has repeated values, then $\bd{P}^T$ has repeated $1$'s in at least a column and  at least a zero column. In other words, assume $rank\left(\bd{P}\right)<N$.
        If we multiply the left of eq. \ref{eq:C2Pn} with $\bd{1}^T$ we get:
       $\bd{1}^T\bd{\Delta}_{N\times\frac{1}{2}\left(N-1\right)N}\bd{c}+\bd{1}^T\bd{n}=\bd{1}^T\bd{P}^T\bd{n}$.
        Because of eq. \ref{eq:Dvanishes}, this reduces to:
       $\bd{1}^T\bd{n}=\bd{1}^T\bd{P}^T\bd{n}$.
       However, for this equation to hold $\bd{P}^T$ must be of full rank. This contradicts the assumption. 
        \end{proof} 
\begin{algorithm}
\caption{Binary algorithm}
\label{alg:binalg}
\begin{algorithmic}[1]
\FOR{$i=0$ to $N-2$}
\FOR{$j=1$ to $N-1$ }
\IF{$x[i]>x[j]$}
\STATE $c[i][j]=1$
\STATE $c[j][i]=0$
\ELSE 
\STATE $c[i][j]=0$
\STATE $c[j][i]=1$
\ENDIF
\ENDFOR
\ENDFOR
\FOR{$(i\neq j)$ and $(j=0 $ to $ N)$}
\STATE $\pi[i]=\pi[i]+c[i][j]$
\ENDFOR
\FOR{$j=0 $ to $ N$}
\STATE $s[\pi[i]]=x[i]$
\ENDFOR
\end{algorithmic}
\end{algorithm}

The cumulative sum is done in parallel too. Since the elements in the sum are binary, it could be possible to use logical circuits to implement this. So it is possible to convert the problem of sorting of N numbers into the problem of N-1 binary cumulative sums. 
For a sequence of length N, the comparison complexity required to sort N numbers ($\mathcal{C}_N$) in terms of the quantity of binary comparisons ($\mathcal{C}_2$) is : 
\begin{equation}\label{eq:bincplx}
\mathcal{C}_{N}\equiv\frac{N^2-N}{2}\mathcal{C}_{2}=\mathcal{O}(N^2)\mathcal{C}_{2}
\end{equation}
Note that we can see this directly from equation \ref{eq:bigC} since there are $\frac{N^2-N}{2}$ disjoint binary comparisons to represent $\bd{C}_N$, so the above equation reads as: A $N$-ary comparison is equivalent to $\frac{N^2-N}{2}\mathcal{C}_{2}$ simultaneous binary comparisons.
 The quadratic term shows that the algorithm is practically useless for large sequences.  Recall that only short sequences are of interest here.
 \begin{figure}[!htb]
\vskip 0.2in
\begin{center}
\includegraphics{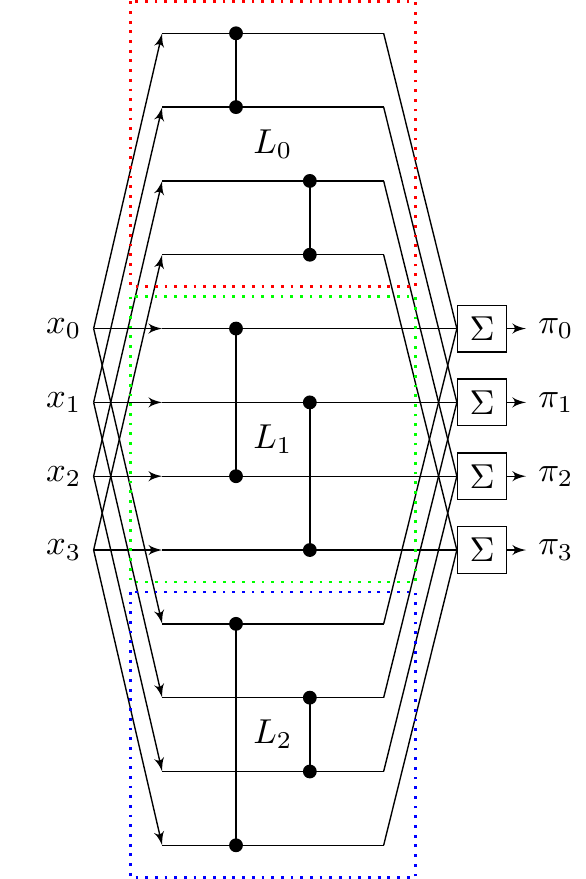}
\caption[Network that yields permutation indices of 4 elements.]{ The comparators only outputs ranks, they don't swap the compared numbers. The ranks from the three parallel levels $L_0$, $L_1$ and $L_2$ are added together in the end. There are three levels and each level has 2 binary comparators.}
\label{fig:net}
\end{center}
\vskip -0.2in
\end{figure}

The question arises if we can minimize the number of parallel levels labelled $L_i$ using ternary or even K-ary comparisons. In fact, this is possible, for example a sequence of length 9 can be separated into 4 parallel nets using ternary comparisons only. The generalization of this for K-ary comparisons is the aim of the next subsection. Furthermore, we would like the levels to be maximal, i.e. the number of K-ary comparators per level is $\frac{N}{K}$.
\FloatBarrier
\subsection{Smallest divisor partitioning of partial ranks} \label{subsec:divisor}
 Let the sequence length $N$ be factored into a pair of divisors:
\begin{equation}
N=d\cdot D
\end{equation}
In the above equation, $d$ is the smallest prime divisor of $N$, and $D$ is the largest divisor of $N$ or the conjugate divisor of $d$ since $D=N/d$.
The topology of the connections of such networks is what determines if the network design sorts correctly. Before stating the general formula, we will enunciate some definitions.

The MATLAB convenient notation for a vector permutation is used, i.e.:
 \begin{equation}\label{eq:MATLABpermutation}
 \bd{y}_{N\times1}=\bd{x}_{N\times1}\left(\bd{\pi}_{N\times1}\right)\Longrightarrow~y_i=x_{\pi_i}; i\in[0,N-1]
 \end{equation}
 The $M\times~L$ matrix $\bd{E}_{i,j,M\times L}$, or simply $\bd{E}_{i,j}$, has 1 in its $\left(i,j\right)$ entry and $0$'s elsewhere. Thus,
  \begin{equation}\label{eq:EeqeeT}
\bd{E}_{i,j}=\bd{\delta}_i\bd{\delta}_{j}^T
\end{equation}
 Let an index vector $\bd{v}_{j,k}\in\mathbb{R} ^{d\times 1}$ be constructed with it's $i$th element defined as:
\begin{equation}\label{eq:indexVectorV}
v_{i,j,k}=(j+k\cdot i)\bmod{D}+D\cdot i
\end{equation}
 where $i\in[0,d-1]$, $j\in[0,D-1]$, $k\in[0,D-1]$.
 Let a second index vector $\bd{w}_{j}\in\mathbb{R} ^{D\times 1}$ with it's $i$th element defined as $w_{i,j}=(i)\bmod{D}+D\cdot j$:
\begin{equation}\label{eq:indexVectorW}
\bd{w}_{j}=\bd{n}_{D\times 1}+i\cdot~D\cdot~\bd{1}_{D\times1}
\end{equation}
 Then we have:
 \begin{equation}\label{eq:CompNeqCompDd}
\bd{\pi}_{N\times1}=\sum_{j=0}^{d-1}\bd{\delta}_{j,d\times 1}\otimes\mathcal{C}_{D\times1}\left(\bd{x}\left(\bd{w}_{j}\right)\right)+\sum_{k=0}^{D-1}\sum_{j=0}^{D-1}\left(\sum_{i=0}^{d-1}\bd{E}_{v_{i,j,k},i,{N\times d}}\right)\mathcal{C}_{d\times1}\left(\bd{x}\left(\bd{v}_{j,k}\right)\right)
\end{equation}
Here $\otimes$ denotes the Kronecker product.
The pattern for the connections are illustrated with an example figure with $N=6$ network that looks like:

\begin{figure}[!htb]
\vskip 0.2in
\begin{center}
\includegraphics{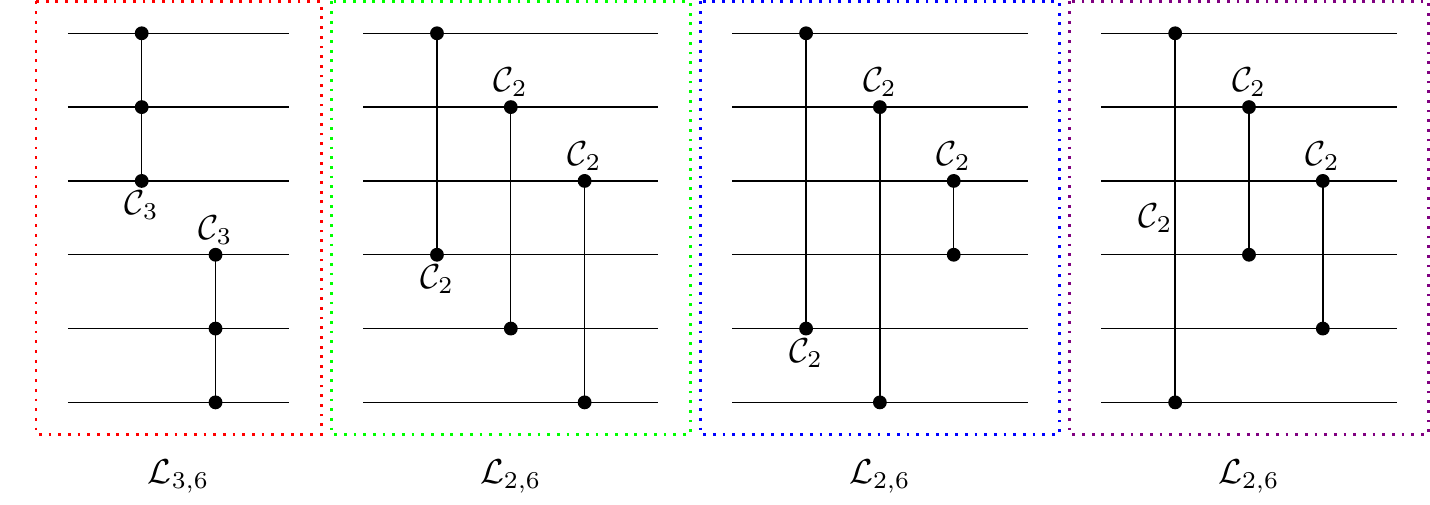}
\caption[Network that yields permutation indices of 6 elements.]{At the left $\mathcal{N}_{3,6}$ is the ternary network, the others are binary networks. The binary and ternary comparators are labeled with $\mathcal{C}_{2}$ and $\mathcal{C}_{3}$ respectively. There are 1 level of ternary comparators and there are 3 levels of binary comparators hence $\mathcal{L}_{6,6}\equiv \mathcal{L}_{3,6}+3\mathcal{L}_{2,6}$. There is 2 ternary comparators in the ternary level and 3 binary ones per binary level, i.e. $\mathcal{C}_{6}\equiv 2\mathcal{C}_{3}+9\mathcal{C}_{2}$.}
\label{fig:net6}
\end{center}
\vskip -0.2in
\end{figure}

We can decompose a sorting network of $N$ numbers into one $D$-ary network and $D$ $d$-ary nets:
\begin{equation}\label{eq:net2net}
\mathcal{L}_{N,N}\equiv 1\cdot\mathcal{L}_{D,N}+D\cdot\mathcal{L}_{d,N}
\end{equation}

Let $\mathcal{C}_{D}$  denote a $D$-ary comparison. Then each network $\mathcal{L}_{d,N}$ contains $D$ $d$-ary comparisons:
\begin{equation}\label{eq:net2comp}
\mathcal{L}_{d,N}\equiv \frac{N}{d}\cdot\mathcal{C}_{d}=D\cdot\mathcal{C}_{d}
\end{equation}

Each comparator unit $\mathcal{C}_{p_i}$ can be modelled by any algorithm, provided it saves the permutation indices. 
 \begin{table}[!htbp]
\begin{minipage}{\textwidth}
  \centering
Partial ranks, or output of $\mathcal{L}_{N,d_j}$ and their sum.
    \begin{tabular}{c c|c c c c c |c}
   &  &\multicolumn{5}{c|}{Partial rank from levels}& \\
  $i$ &$x_i$&$\mathcal{L}_{8,4}$&$\mathcal{L}_{8,2}$&$\mathcal{L}_{8,2}$&$\mathcal{L}_{8,2}$&$\mathcal{L}_{8,2}$&$\pi_i$\\ \hline
0&5  &2 &0&0&0&0&2\\
1&12&3&1&1&1&1&7\\
2&2  &0&0&0&0&0&0\\
3& 3 &1&0&0&0&0&1\\
4&5  &0&1&1&1&0&3\\
5&7 &0&1&1&1&0&5\\
6&8 &3&1&0&1&1&6\\
7&6&1&1&1&0&1&4\\
\end{tabular}
  \caption[Example Divisors.]{Example of summing partial ranks with divisor partitioning (eq.\ref{eq:CompNeqCompDd}) with $N=8$. Here, $D=4$ and the first column of outputs corresponds to 2 quaternary comparator ranks placed one above the other, hence it's partial ranks go from $0$ to $3$. The last column are the ranks, or the sum of all the partial ranks.}\label{tab:ExampleD}%
\end{minipage}
\end{table}

Substituting eq.\ref{eq:net2comp} into \ref{eq:net2net} we get the mixed comparisons complexity:
\begin{equation}\label{eq:compdiv}
\mathcal{C}_{N}\equiv d\cdot\mathcal{C}_{N/d}+\left(\frac{N}{d}\right)^2\cdot\mathcal{C}_{d}
\end{equation}
If we take take eq. \ref{eq:compdiv}, and substitute in eq.\ref{eq:bincplx}, we show that the binary complexity are the same for both algorithms:
\begin{equation}\label{eq:divbinequivalentnumberofcomp}
\mathcal{C}_{N}\equiv d\cdot\frac{1}{2}\left(\left(\frac{N}{d}\right)^2-\frac{N}{d}\right)\mathcal{C}_{2}+\left(\frac{N}{d}\right)^2\cdot\frac{1}{2}\left(d^2-d\right)\mathcal{C}_{2}=\frac{1}{2}\left(N-1\right)N\mathcal{C}_{2}
\end{equation}
The above relation is a hint that both algorithms are equivalent.
If we use algorithm \ref{alg:binalg}, to model all comparisons, the whole network reduces to binary comparisons. This effectively reduces the conjugate divisor algorithm into alg. \ref{alg:binalg}. Although this substitution wouldn't be interesting because the goal of this section was to see if it was possible to use non binary comparators, this reduction means that the two algorithms are equivalent.

\begin{proof}

 If we take eq.\ref{eq:CompNeqCompDd} and substitute all comparisons with eq.\ref{eq:permutationvector} without summing in the columns together we get:
 \begin{equation}\label{eq:CNeqCDd}
\bd{C}_{N}=\sum_{j=0}^{d-1}\bd{E}_{j,j,d\times d}\otimes\bd{C}_{D\times D}\left(\bd{x}\left(\bd{w}_{j}\right)\right)+\sum_{k=0}^{D-1}\sum_{j=0}^{D-1}\bd{C}_{d\times d}\left(\bd{x}\left(\bd{v}_{j,k}\right)\right)\otimes\bd{E}_{k,j,D\times D}
\end{equation}
The above equation shows that the binary matrix $\bd{C}_{N}$ can be represented as an addition of small discontinuous tiles i.e., $\bd{C}_{D}$'s and $\bd{C}_{d}$'s.
\end{proof}
\FloatBarrier
\subsection{Prime partitioning of partial ranks} \label{subsec:factor}
Let 
\begin{equation}
N=\prod_{i}^m f_i=\prod_{j}^n p_j^{k_j}
\end{equation}
Where $f_i$ are prime factors of $N$ and $f_1\le\cdots\le f_m$. The $p_j$'s are the unique prime factors and $k_j$ is their corresponding power and $p_1<\cdots<p_n$.
If $D$ can be further decomposed into prime factors then we can reapply the algorithm iteratively on equation \ref{eq:net2net} until no further decompositions are possible. Using this iterative scheme, we can sort $N$ numbers using minimal sized partitions consisting of the prime factors of $N$. We start the steps by rewriting equation ~\ref{eq:net2net}:
\begin{equation}\label{eq:net2netp}
\mathcal{L}_{N,N}\equiv 1\cdot\mathcal{L}_{\frac{N}{f_1},N}+\frac{N}{f_1}\cdot\mathcal{L}_{f_1,N}
\end{equation}
We then repeat the recursion until we can't factor any more:

$\mathcal{L}_{N,N}\equiv \mathcal{L}_{\frac{N}{f_1},N}+\frac{N}{f_1}\cdot\mathcal{L}_{f_1,N}\equiv (\mathcal{L}_{\frac{N}{f_{1}f_{2}},N}+\frac{N}{f_{1}f_{2}}\cdot\mathcal{L}_{f_2,N})+\frac{N}{f_1}\cdot\mathcal{L}_{f_1,N}\equiv \cdots$

We then end up with the sum:
\begin{equation}\label{eq:netP}
\mathcal{L}_{N,N}\equiv \sum_{i=1}^m \frac{N}{\prod_{j=1}^i f_j}\mathcal{L}_{f_i,N}=\sum_{i=1}^m \left(\prod_{j=i}^m f_j \right)\frac{\mathcal{L}_{f_i,N}}{f_i}
\end{equation}
This sequence can be then be organised wrt to it's unique prime factors:
\begin{equation}\label{eq:netP2}
\mathcal{L}_{N,N}\equiv \sum_{i=1}^n \frac{N}{\prod_{j=1}^i p_j^{k_j}}\frac{p_i^{k_i}-1}{p_i-1}\mathcal{L}_{p_i,N}
\end{equation}
Applying equation \ref{eq:net2comp} to \ref{eq:netP}, we get a formula for the different types of comparisons:
\begin{equation}\label{eq:compP}
\mathcal{C}_{N}\equiv \sum_{i=1}^m \frac{N}{\prod_{j=1}^i f_j}\left(\frac{N}{f_i}\right)\mathcal{C}_{f_i}=N\sum_{i=1}^m \left(\prod_{j=i}^m f_j \right)\frac{\mathcal{C}_{f_i}}{f_i^2}
\end{equation}
\begin{equation}\label{eq:netP3}
\mathcal{C}_{N}\equiv \sum_{i=1}^n \frac{N}{\prod_{j=1}^i p_j^{k_j}}\frac{p_i^{k_i}-1}{p_i-1}\frac{N}{p_i}\mathcal{C}_{p_i}
\end{equation}
The proof for eq.\ref{eq:netP3} follows the same lines as the proof for equation \ref{eq:nPpow}. If we substitute each $\mathcal{L}_{p_i,N}$ in equation ~\ref{eq:netP} with 1, we get the number of column vectors of keys to be added in the final adder step:
\begin{equation}\label{eq:addsP}
|\mathcal{L}_N|=\sum_{i=1}^m \frac{N}{\prod_{j=1}^i f_j} 
\end{equation}
Consequently, the addition complexity $\mathcal{C}_{N}^{\bigoplus}=|\mathcal{L}_N|-1$.
%=\sum_{i=1}^m \frac{\prod_{j=i}^m p_j}{p_i} 
A special case of eq. \ref{eq:netP} when $N=p^k$ include:
\begin{equation}\label{eq:nPpow}
\mathcal{L}_{N,N}\equiv \frac{N-1}{p-1}\mathcal{L}_{p,N}
\end{equation}
\begin{proof}
We start with equation \ref{eq:netP} and set all primes $p_i=p$.
We get a geometrical series $\mathcal{L}_{N,N}\equiv \sum_{i=1}^m p^{m-i}\mathcal{L}_{p,N}=p^{m-1}\frac{1-p^{-m}}{1-p^{-1}}\mathcal{L}_{p,N}$ and \ref{eq:nPpow} follows.
\end{proof}
The comparison complexity for prime powers is:
\begin{equation}\label{eq:cPpow}
\mathcal{C}_{N}\equiv \frac{N-1}{p-1}\cdot\frac{N}{p}\mathcal{C}_{p}
\end{equation}
A second case of eq. \ref{eq:netP} when $N=p_1^{k_1}p_2^{k_2}$ include:
\begin{equation}\label{eq:nPpow2}
\mathcal{L}_{N,N}\equiv p_2^{k_2}\frac{p_1^{k_1}-1}{p_1-1}\mathcal{L}_{p_1,N}+\frac{p_2^{k_2}-1}{p_2-1}\mathcal{L}_{p_2,N}.
\end{equation}
Now upon inspecting the formula for $|\mathcal{L}_N|$ in equation \ref{eq:addsP} we see that the upper bound is obtained when the number of prime factors $m$ is maximal and the denominator is minimal. This happens only when $N=2^m$. Then $|\mathcal{L}_{2^m}|=N-1$. The minimum is when $N$ is prime and hence $|\mathcal{L}_{p}|=1$
 and so we have:
\begin{equation}\label{eq:addspow2}
1\leq|\mathcal{L}_{N}|\leq N-1 ; N\geq 2
\end{equation}
The formula provided in the sequence of OEIS \textbf{A006022} \cite{addprime}, (related to the Maundy cake problem \cite{berlekampwinning}) is:
\begin{equation}\label{eq:addspow2reference}
a\left(N\right)=max\left(\left[d\cdot a\left(N/d\right)+1\right]_{\forall d|N}\right)
\end{equation}
applying the iteration until we cover all the values we end up with the formula for $|\mathcal{L}_{N}|$. In other words, $a\left(N\right)=|\mathcal{L}_{N}|$ implies that $|\mathcal{L}_{N}|$ are Nim numbers.

If we are interested in the total number of comparisons regardless of the number of input the comparators take we get by setting all $\mathcal{C}_{f_i}$ to $1$:
\begin{equation}\label{eq:comptotal}
|\mathcal{C}_{N}|\equiv \sum_{i=1}^m \frac{N^2}{f_i\prod_{j=1}^i f_j}
\end{equation}
It can be shown using the same arguments to show the limits in \ref{eq:addspow2} that:
\begin{equation}\label{eq:compspow2}
1\leq|\mathcal{C}_{N}|\leq \frac{N^2-N}{2} ; N\geq 2
\end{equation}
The first $50$ terms are:
\[0,1,1,6,1,11,1,28,12,27,1,58,1,51,28,120,1,105,1,154,52,123,1,260,30,171,117,298,\]
\[1,281,1,496,124,291,54,534,1,363,172,708,1,545,1,730,309,531,1,1096,56,685.\]
 \begin{table}[!htbp]
\begin{minipage}{\textwidth}
  \centering
    \begin{tabular}{c|c c c c c c c |c}
   &  \multicolumn{7}{c|}{$p$}& \\
  $N$ &2&3&5&7&11&13&$\dots$&$|\mathcal{L}_{N}|$\\ \hline
2&1  & & & & & & &1\\
3&   &1& & & & & &1\\
4&3  && & & & & &3\\
5&    & &1& & & & &1\\
6&3  &1& & & & & &4\\
7&    & & &1& & & &1\\
8&7 & & & & & & &7\\
9&    &4& && & & &4\\
10&5 & &1&& & & &6\\
11&& &&&1& & &1\\
12&9&1&&&& & &10\\
13&&&&&&1& &1\\
14&7&&&1&& & &8\\
15&&5&1&&& & &6\\
16&15&&&&& & &15\\
\end{tabular}
  \caption[Coefficients for number of layers.]{Coefficients in $\mathcal{L}_{p,N}$ of eq.\ref{eq:netP2} and  $|\mathcal{L}_{N,N}|$. }\label{tab:Ls}%
\end{minipage}
\end{table}

 \begin{table}[!htbp]
\begin{minipage}{\textwidth}
  \centering
    \begin{tabular}{c|c c c c c c c |c}
   &  \multicolumn{7}{c|}{$p$}& \\
  $N$ &2&3&5&7&11&13&$\dots$&$|\mathcal{C}_{N}|$\\ \hline
2&1  & & & & & & &1\\
3&   &1& & & & & &1\\
4&6  && & & & & &6\\
5&    & &1& & & & &1\\
6&9  &2& & & & & &11\\
7&    & & &1& & & &1\\
8&28 & & && & & &28\\
9&    &12& && & & &12\\
10&25 & &2&& & & &27\\
11&& &&&1& & &1\\
12&54&4&&&& & &58\\
13&&&&&&1& &1\\
14&49&&&2&& & &51\\
15&&25&3&&& & &28\\
16&120&&&&& & &120\\
\end{tabular}
  \caption[Coefficients for number of comparisons.]{Coefficients in $\mathcal{C}_{p_i}$ of eq.\ref{eq:netP3} and  $|\mathcal{C}_{N}|$. }\label{tab:Cs}%
\end{minipage}
\end{table}

\FloatBarrier
\bibliography{bibliography}
\bibliographystyle{plain}
\end{document}